\documentclass[12pt,preprint]{aastex}

\newcommand{\kev}{\:{\rm ke\!V}}

\newcommand{\ergl}{ergs~s$^{-1}$}

\newcommand{\ergcms}{erg~cm$^{-2}$~s$^{-1}$}

\newcommand{\cps}{${\rm ct}\, {\rm s}^{-1}$}

\newcommand{\chisq}{$\chi^2$}

\newcommand{\CXO}{{\sl Chandra X-ray Observatory}}
\newcommand{\chandra}{{\sl Chandra}}
\newcommand{\ROSAT}{{\sl ROSAT}}

\def\ha{H$\alpha$}
\def\hb{H$\beta$}
\def\hz{H$\zeta$}
\def\hg{H$\gamma$}
\def\he{H$\epsilon$}
\def\hd{H$\delta$}

\def\oiii{[O~{\sc iii}]}
\def\oii{[O~{\sc ii}]}
\def\nii{[N~{\sc ii}]}
\def\sii{[S~{\sc ii}]}

\def\l{$\lambda$}
\def\oi{[O~{\sc i}]}
\def\neiii{[Ne~{\sc iii}]}
\def\nev{[Ne~{\sc v}]}
\def\fevii{[Fe~{\sc vii}]}
\def\fex{[Fe~{\sc x}]}
\def\feii{Fe~{\sc ii}}

\def\hei{He~{\sc i}}

\def\ltsim{\mathrel{\hbox{\rlap{\hbox{\lower4pt\hbox{$\sim$}}}\hbox{$<$}}}}

\slugcomment{2005.08.30, \apj \,  resubmitted}
\shorttitle{{The First \chandra\ Field}}
\shortauthors{}

\begin{document}

\title{{The First \chandra\ Field\footnote{Based in part on observations made at the European Southern Observatory, La Silla, Chile}}}

\author{
Martin C. Weisskopf,\altaffilmark{2}
Thomas L. Aldcroft,\altaffilmark{3}
Robert A. Cameron,\altaffilmark{{3,4}}\\
Poshak Gandhi,\altaffilmark{5}
C\'edric Foellmi,\altaffilmark{5}
Ronald F. Elsner,\altaffilmark{2}
Sandeep K. Patel,\altaffilmark{6}\\
Kinwah Wu,\altaffilmark{7}
and Stephen L. O'Dell,\altaffilmark{2}
}

\altaffiltext{2}
{NASA Marshall Space Flight Center, NSSTC XD12, 320 Sparkman Drive, Huntsville, AL 35805.}
\altaffiltext{3}
{Harvard--Smithsonian Center for Astrophysics, 60 Garden Street, Cambridge, MA 02138.}
\altaffiltext{4}
{Stanford Linear Accelerator Center, SLAC MS-43A, 2575 Sand Hill Road, Menlo Park, CA 94025.} 
\altaffiltext{5}
{European Southern Observatory, Alonso de Cordova 3107, Vitacura, Casilla 19001, Santiago, Chile.}
\altaffiltext{6}
{Universities Space Research Association, NSSTC XD12, 320 Sparkman Drive, Huntsville, AL 35805.}
\altaffiltext{7}
{Mullard Space Science Laboratory, University College London, Holmbury St.\ Mary, Dorking, Surrey RH5 6NT, United Kingdom.}

\begin{abstract}

Before the official first-light images, the \CXO\ obtained an X-ray image of the  field to which its focal plane was first exposed.
We describe this historic observation and report our study of the first \chandra\ field.
\chandra's Advanced CCD Imaging Spectrometer (ACIS) detected 15 X-ray sources, the brightest being dubbed ``Leon X-1'' to honor the \chandra\ Telescope Scientist, Leon Van Speybroeck.
Based upon our analysis of the X-ray data and spectroscopy at the European Southern Observatory (ESO; La Silla, Chile), we find that Leon X-1 is a Type-1 (unobscured) active galactic nucleus (AGN) at a redshift $z=0.3207$. 
Leon X-1 exhibits strong \feii\ emission and a broad-line Balmer decrement that is unusually flat for an AGN.
Within the context of the Eigenvector-1 correlation space, these properties suggest that Leon X-1 may be a massive ($\ge 10^{9}\,M_{\odot}$) black hole, accreting at a rate approaching its Eddington limit.

\end{abstract}


\section{Introduction}

On 1999 July 23 00:31 EDT (1999:204:04:31 UTC), after attempts the evenings of July 19 and July 21, the Space Shuttle {\sl Columbia} mission STS-93 launched from NASA's Kennedy Space Center.  
A little over 7 hours later, {\sl Columbia} deployed the {\CXO}. 
(For a detailed description of the Observatory and its instruments see e.g. Weisskopf et al. (2005) and references therein.)
About an hour thereafter, two firings of the attached solid-rocket Inertial Upper Stage and subsequent separation sent \chandra\ toward a high elliptical orbit.
Five burns of \chandra's Integral Propulsion System---on July 24, 25, and 27 and August 4 and 7---placed the \CXO\ into its operational orbit, with initial apogee and perigee altitudes of 140 Mm and 10 Mm, respectively.

During the orbital-transfer period, the science-instrument teams activated and began functional checks of the High-Resolution Camera (HRC) and of the Advanced CCD Imaging Spectrometer (ACIS).  
Also during this period (on July 26), bake-out of the ACIS began, with its door to the optical cavity of the telescope still closed so that any molecular contamination would vent to space.
After approximately two weeks (on August 8), the \chandra\ Operations Control Center (OCC) commanded open the ACIS door.
Then, on 1999 August 12, the OCC commanded open the telescope's (forward contamination cover) sunshade door, the last barrier between the \chandra\ focal plane and the X-ray sky.

When the sunshade door opened, the attitude control system was not yet in its normal operating mode. 
Although the aspect camera was acquiring data, it was not yet an active part of the attitude control system. 
Rather, the {\sl Chandra} gyroscopes, which have very small drift, were controlling the spacecraft's attitude.
Further, the dither mode, which moves the image around in a small focal-plane region (to avoid focusing onto a single pixel) was not yet active.
Consequently, the pointing was already very stable, even without aspect-camera data.  
However, the absolute pointing was uncertain by up to 10 degrees, because the fine attitude of the spacecraft had not yet been established. 

As \chandra's first celestial X rays reflected off its precision mirrors, the ACIS-S (spectroscopy array) lay in the focal position, with the telescope's aim point on CCD S3---one of two back-illuminated devices of the 10 CCDs comprising the ACIS focal plane.
The \chandra\ science team had selected this as the at-launch configuration, in case the translation table failed to operate.
Likewise, the at-launch configuration of the focus mechanism placed S3 near best focus, based upon x-ray testing at MSFC's X-Ray Calibration Facility (XRCF).

With X rays falling onto the active ACIS detector, the \chandra\ science team obtained the Observatory's first image---a 9-ks observation using ACIS CCDs S2, S3, I0, I1, I2, and I3. 
It is impossible to convey the excitement and tension that accompanied this observation, as the image of the first photons appeared on a display at the OCC. 
The \chandra\ Project, which had its formal origins in an unsolicited 1976 proposal (Riccardo Giacconi, Principal Investigator; Harvey Tananbaum, Co-Principal Investigator) to NASA, was 23 years in the making.
Everyone present at the OCC keenly felt the exhilaration of witnessing these efforts come to fruition.

\begin{figure}
\begin{center}
\includegraphics[angle=-90]{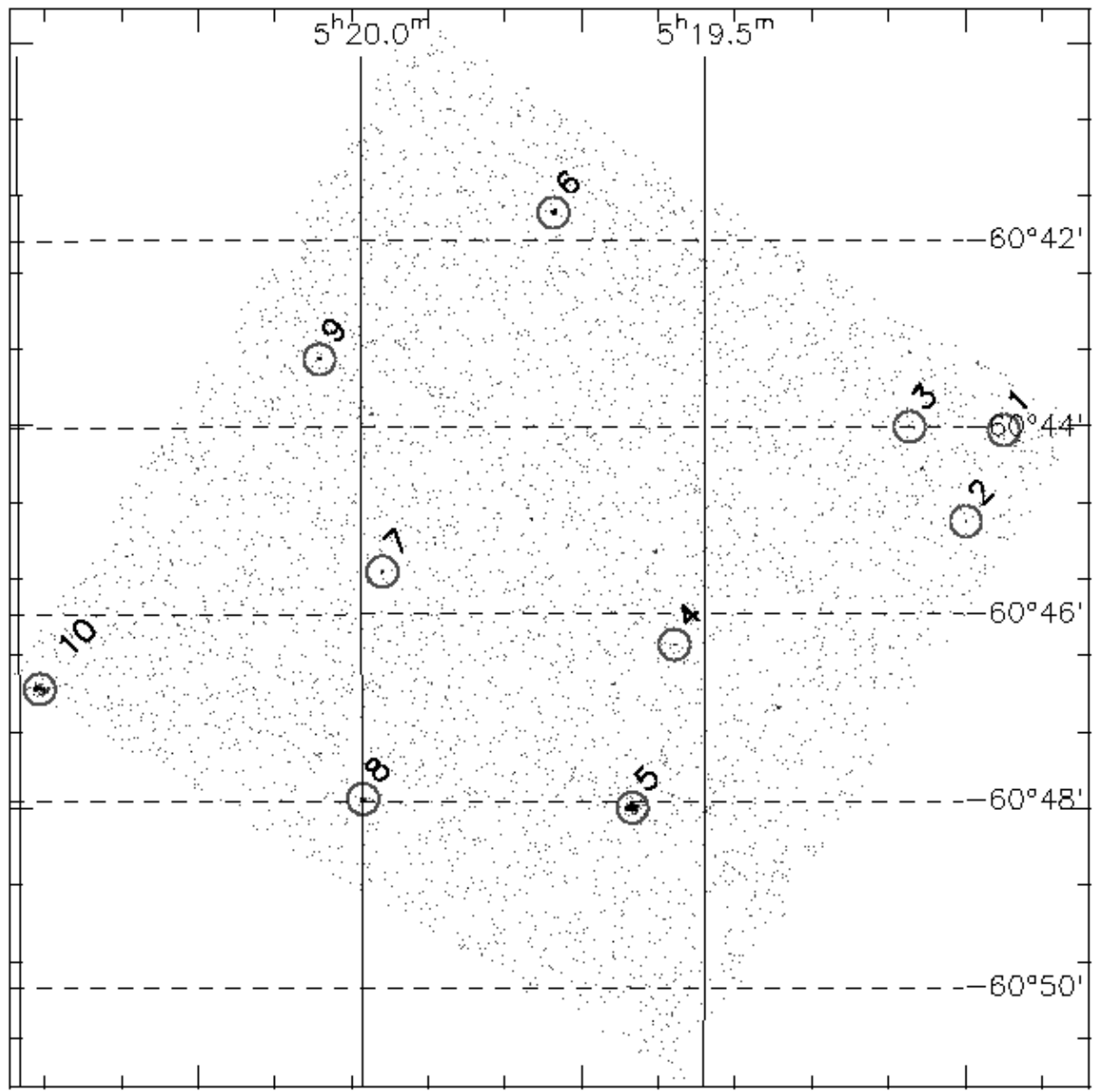}
\end{center}
\caption{X-ray image of the first \chandra\ field, on the ACIS S3 (back-illuminated) CCD. 
A 5\arcsec-radius circle encloses each of the 10 sources detected on S3.
The brightest source---S3-5, dubbed ``Leon X-1''---is an AGN at $z=0.3207$.
Although obtained prior to on-orbit focus adjustment, the image demonstrated the unprecedented angular resolution of \chandra.
\label{f:s3-image}}
\end{figure}

After several minutes, the accumulated ACIS image (Figure~\ref{f:s3-image}) showed a few photons concentrated within a few arcseconds of each other, not too far (about 5 arcminutes) from the optical axis. 
This immediately demonstrated that \chandra's mirrors were performing as expected and that the telescope was not far from best focus.
The clustering of photons was from the brightest source in \chandra's first field.
To acknowledge the enormous contributions to \chandra\ by the Telescope Scientist Leon Van Speybroeck, the Project Scientist (MCW) dubbed this first source ``Leon X-1''.
\clearpage
\clearpage
The \chandra\ team next pointed the Observatory toward the radio-loud quasar PKS 0637-752 ($z=0.654$), selected for on-orbit check-out and imaging optimization.
After bore-sighting and focusing the telescope, the \chandra\ ACIS obtained the first publicly released image on August 15, showing the discovery of an X-ray jet in this quasar (Schwartz et al.\ 2000).
On August 19, \chandra\ obtained the {\em official} ``first-light'' image (released on August 26; Tananbaum (1999); Pavlov et al. (2000))---a spectacular view of the supernova remnant Cassiopeia A (Cas A), revealing a central, candidate neutron star.

Here we present an analysis of those data obtained during the true ``first light''---the image of the first \chandra\ field and its brightest source, ``Leon X-1''.
First (\S \ref{s:obs}) we describe the \chandra\ observation and analyses of the X-ray data.
Next (\S \ref{s:vis_sp}) we present and discuss visible-light spectroscopy of ``Leon X-1'' obtained at the European Southern Observatory.
We conclude (\S \ref{s:agn_prop}) with a discussion of the multi-wavelength properties of ``Leon X-1'' in the context of the E1 correlation space for Active Galactic Nuclei (AGNs).

\section{\chandra\ Observation and Data Analysis \label{s:obs}}

The first {\sl Chandra} observation (ObsID 62568) lasted 9 ks and used ACIS CCDs S2, S3, I0, I1, I2, I3 in the timed-exposure mode, with the standard 3.241-s frame time. 
During the observation, the aspect camera tracked 6 stars that it had 
acquired during a full-field search.  
This was unique in that the aspect stars were selected autonomously, rather than by ground command.
In addition, the initial on-board attitude estimate differed by approximately 7 degrees from the true attitude.  
Consequently, we prepared special input files for a custom run of the aspect data-processing pipeline to produce an aspect solution.
This solution showed that the Observatory was pointed at approximately  $\alpha_{2000} = 5^{\rm h}\,19^{\rm m}\,34^{\rm s}$ and $\delta_{2000} = -60\degr\, 43\arcmin\, 11\arcsec$. 
However, this position was still uncertain by several arcseconds because the
observation was performed without tracking the aspect fiducial lights, which enable
precise registration of the science-instrument focal plane relative to the
telescope boresight.  
Hence, we fine-tuned the aspect solution (\S \ref{s:analysis_i}) using accurately known positions of optical counterparts obviously identified with  X-ray sources in the field.  

We used \chandra\ X-ray Center (CXC) processing (CXCDS 6.2.4) to create  
Level-2 event lists. 
For finding X-ray sources, we selected events in pulse-invariant (PI) channels corresponding to 0.5--8.0\,$\kev$ for the front-illuminated CCDs (S2, I0--I3), 
and to 0.25--8.0\,$\kev$ for the back-illuminated CCD (S3).  
Due to uncertainties in the low-energy response, we used only data in the range 0.5--8.0\,$\kev$ for the spectral analyses.  
During the observation, there were no instances of increased background. 
Following a discussion (\S \ref{s:analysis_i}) of our analysis of the X-ray image, we describe (\S \ref{s:analysis_s}) the results of X-ray spectrometry of these sources and then summarize (\S \ref{s:xray_desc}) their X-ray properties.

\subsection{Image Analysis \label{s:analysis_i}}

We employed source-finding techniques described in Swartz et al.~(2003),
with a circular-Gaussian approximation to the point spread function and a
minimum signal-to-noise ratio (S/N) of 2.6, resulting in much fewer than one expected accidental detection in the field. 
The corresponding background-subtracted point-source detection limit is about 10 counts, corresponding to a 0.5--8.0\,-$\kev$ flux of about $7\!\times\! 10^{-15}$ \ergcms\ for an unabsorbed power law of photon index $\Gamma=1.5$.  
The algorithm found 15 sources---1 on S2, 10 on S3  (Figure~\ref{f:s3-image}), and 4 on the (4-CCD) I array. 
That most of the detected sources were on S3 is not surprising, in that this (back-illuminated) CCD included the aim point and also has the deepest sensitivity of the ACIS CCDs because of its superior low-energy response.

\subsubsection{X-ray Source Positions \label{s:positions}}

After detection of the sources, we immediately noticed a systematic offset of
about 5\arcsec\ between several of the X-ray sources and their respective candidate visible-light counterparts in the United States Naval Observatory Catalog USNO-B1.0 (Monet et al.~2003; hereafter, USNO-B1), one being a $7^{\rm th}$-magnitude star also detected with the aspect camera. 
In order to fine tune the aspect solution, we minimized the separation between the X-ray and visible-light positions (\S \ref{s:counterparts}) using a position-error-weighted least-squares fit, treating the right ascension, declination, and roll angle of the pointing position as free parameters. 
For the visible-light positions, we adopted uncertainties from the USNO-B1 catalog. 
For the X-ray positions, we used uncertainties given by  $1.51(\sigma^2/N+{\sigma_s}^2)^{1/2}$, where $\sigma$ determines the size of the circular Gaussian that approximates the point spread function at the source location, $N$ is the aperture-corrected number of source counts, $\sigma_s$ is a systematic error, and the factor 1.51 scales the radius to enclose 68\% of the circular Gaussian.  
For various values of $\sigma_s$ ranging from $0\farcs0$ to $0\farcs4$, we allowed $\delta$RA, $\delta$Dec, and $\delta$Roll to vary freely. 
Once we applied the resulting offsets to the initial X-ray positions, all fits were excellent, independent of the choice for $\sigma_s$. 
Uncertainties in the plate scale\footnote{see
http://asc.harvard.edu/cal/Hrma/optaxis/platescale/} imply a systematic 
uncertainty of $0\farcs13$.
Given that, where relevant, the USNO and \chandra\ positions agree to high precision, we believe that $0\farcs2$ is a reasonable and conservative estimate for $\sigma_s$. 
We note that the precise value of $\sigma_s$ has no statistically significant impact on the positions of the X-ray sources. 

Table~\ref{t:xsources} lists positions of the 15 detected X-ray sources, with corresponding extraction radius, net counts, signal-to-noise ratio, and X-ray positional uncertainty. 
The field's brightest X-ray source, S3-5, is the source we nicknamed ``Leon X-1''.
Table~\ref{t:xsources} also gives the X-ray flux and the separation between each X-ray source and any candidate counterpart (\S \ref{s:counterparts}) in the USNO-B1 or in the Two-Micron All Sky Survey (2MASS).

\subsubsection{Search for Counterparts\label{s:counterparts}}

Using the HEASARC {\tt cfeature},\footnote{see
http://heasarc.gsfc.nasa.gov/db-perl/W3Browse/w3browse.pl} we searched available catalogs for candidate counterparts centered on the X-ray positions listed in
Table~\ref{t:xsources}. 
We selected non-X-ray candidate counterparts by searching around the X-ray position within a 99\%-confidence radius ($r_{99}$, 3.03/1.51 times the X-ray positional uncertainty $\sigma_{X}$ listed in column 6 of Table~\ref{t:xsources}).
Although approximating the point-spread function by a circular Gaussian becomes less accurate far off-axis, use of this approximation has virtually no impact on the coordinates of the X-ray sources. 

\paragraph{USNO\label{s:USNO}}
Table~\ref{t:counterparts} lists the positions of  candidate visible-light counterparts (with two candidates for source I-3).
There are 2446 USNO-B1.0 sources within a 12\arcmin\ radius centered on the X-ray pointing direction, corresponding to $1.5 \times 10^{-3}$ USNO sources arcsec$^{-2}$.
Based upon this sky density, column 4 of Table~\ref{t:counterparts} gives the expected number $N_{99}$ of accidental USNO coincidences within $r_{99}$ of each X-ray source. 
The probability of getting one or more matches by chance is given by the Poisson probability, $(1-e^{-N_{99}}) \rightarrow N_{99}$ for $N_{99}\ll 1$. 
In most cases (9 of 16), these probabilities are below 1\%.
For the two X-ray sources most off axis---IA-2 and IA-4---the probability for accidental identification exceeded 10\%.
Consequently, we chose not to search for USNO counterparts for these two sources. 
Column 8 of Table~\ref{t:xsources} gives the separation between each X-ray source and its USNO candidate, where we found one.
We note that the candidate optical counterpart to source S3-10 is a $7^{\rm th}$-magnitude A3 V star. 
Due to its brilliance, other viable counterpart candidates may be hidden (see \S \ref{s:S3-10}).

\paragraph{2MASS\label{s:2MASS}}
We found (Table~\ref{t:counterparts}) 3 candidate 2MASS counterparts.
There are 6427 2MASS sources within a 12\arcmin\ radius centered on the X-ray pointing direction, corresponding to $3.9 \times 10^{-3}$ 2MASS sources arcsec$^{-2}$.
Based upon this sky density, column 7 of Table~\ref{t:counterparts} gives the $N_{99}$ of accidental 2MASS coincidences within $r_{99}$ of each X-ray source. 
Similar to our search of the USNO catalog, we chose not to search for 2MASS counterparts to the four X-ray sources most off-axis---IA-2, IA-3, and IA-4.

Tables~\ref{t:xsources} and \ref{t:counterparts} list additional pertinent information about potential infrared counterparts. 
Column 9 of Table~\ref{t:xsources} gives the separation between each X-ray source and its 2MASS candidate, where we find one.
Column 8 of Table~\ref{t:counterparts} gives the separation between visible-light and infrared candidates for those X-ray sources having both USNO and 2MASS candidates.
In these three cases, USNO--2MASS separations are subarcsecond, indicating that the visible-light and infrared sources are the same object. 

\begin{figure}
\begin{center}
\includegraphics[scale=.7,angle=-90]{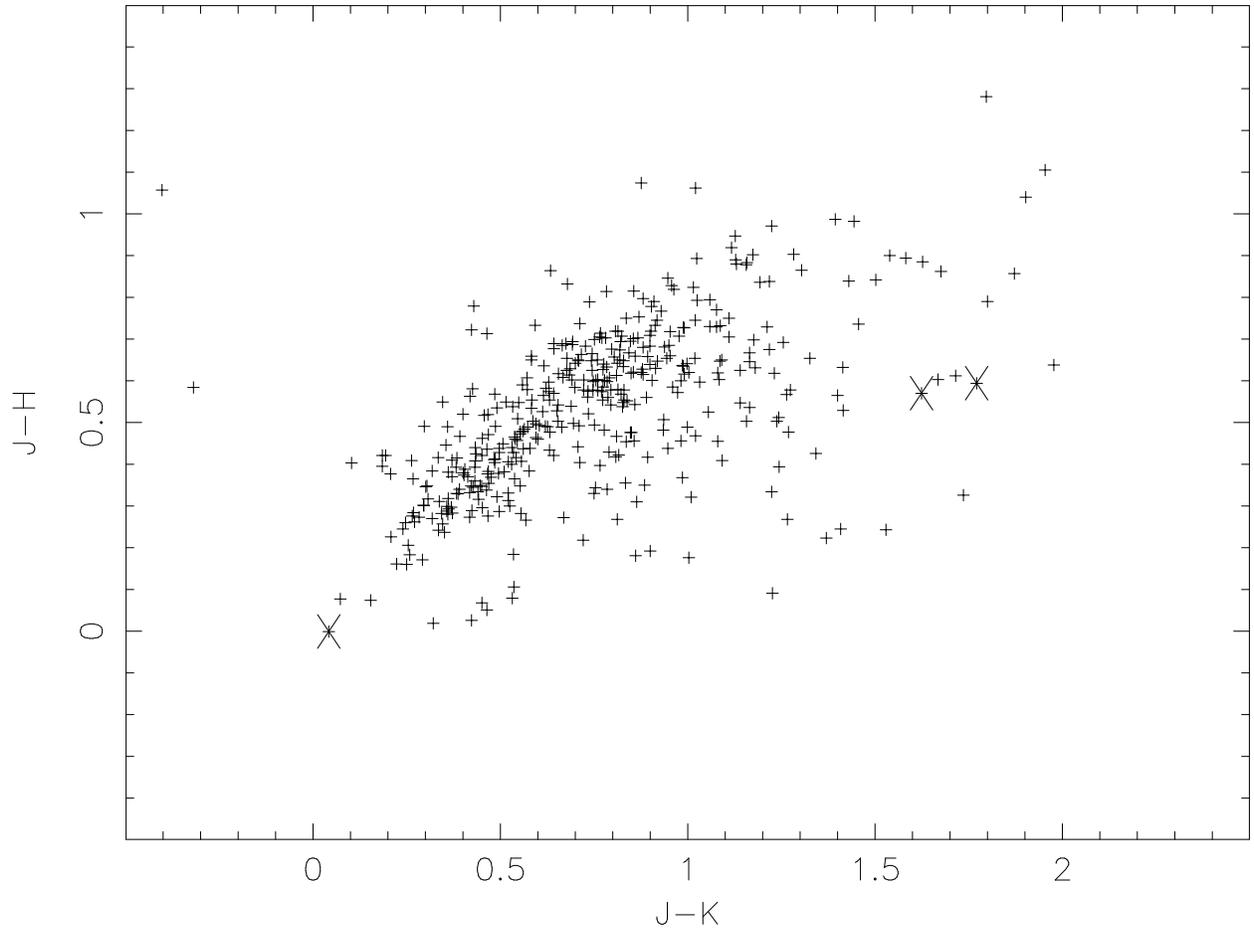}
\end{center}
\caption{Near-infrared color--color diagram for 2MASS field stars located within a 12\arcmin\ radius from the center of the \chandra\ field. 
An ``X'' denotes a 2MASS counterpart to a \chandra\ X-ray source---from left to right, S3-10 (an A3 V star), S3-9 (probable AGN), S3-5 (confirmed AGN ``Leon X-1'').
\label{f:2MASS_colors}}
\end{figure}

Table~\ref{t:2MASS_colors} gives the near-infrared magnitudes and colors of the three 2MASS candidate counterparts.
Figure~\ref{f:2MASS_colors} shows a near-infrared color--color diagram of 2MASS objects within the \chandra\ field, including the 3 candidate counterparts. 
The 2MASS colors for the A3 V star (Table~\ref{t:2MASS_colors})---the counterpart to S3-10---are consistent with its stellar classification. 
However, the other two 2MASS candidate counterparts do not lie on the main sequence, independent of the amount of absorption (reddening).
These two objects---counterparts to S3-5 (Leon X-1) and to S3-9---have similar infrared colors, consistent with those of an unobscured AGN. 
Indeed, spectroscopy of Leon X-1 (\S \ref{s:vis_sp_obs}) at the European Southern Observatory (ESO) confirms that it is a Type-1 AGN at $z=0.3207$.
\clearpage
\clearpage
\paragraph{Other X-Ray Observations\label{s:other_x-ray}}

We also used the HEASARC ``Browse''  feature to search for other X-ray observations of the \chandra\ sources. 
Only various \ROSAT\ catalogs yielded positional coincidences within 0.5\arcmin\  radius: 
The possible faint ($(9.9\pm 3.9)\times 10^{-3}$ \cps) source 1RXS~J052145.5-602906 lies about $14\arcsec$ from IA-4; the bright ($(6\pm 1)\times 10^{-2}$ \cps)) source 1RXS~J051934.0-604800, about $16\arcsec$ from Leon X-1. 
Owing to uncertainties in the \ROSAT\ positions, these separations are consistent with the respective \ROSAT\ and \chandra\ sources being the same in each case. 

\subsection{X-ray Spectral Analysis\label{s:analysis_s}}

For each source for which we performed a detailed spectral analysis we obtained source counts and then spectrum from within the appropriate extraction radius (Table~\ref{t:xsources}).  
To estimate background for the spectral analysis discussed here, we created one data set each for S2, S3, and the I array, removing all events within 10 times the respective extraction radius of each source, and inferred the average backround per square arcsecond. 
For S3 the rate was $2.2\times 10^{-6}$ \cps~sq-arcsec$^{-1}$. 

For the spectral analysis, we employed CIAO 3.0.2 to extract the pulse-invariant (PI) files and used CXC CALDB 2.25 calibration files (gain maps, quantum-efficiency uniformity, and effective area) to generate effective-area and response functions.
We grouped the data to ensure at least 15 counts per spectral bin and restricted the energy range to 0.5--8.0~$\kev$, due to uncertainties in the ACIS spectral response at lower energies.
In calculating interstellar absorption, we utilized {\tt tbabs} (available in XSPEC \footnote{see http://heasarc.gsfc.nasa.gov/docs/xanadu/xspec/} v.11.2) with default abundances and cross sections (Wilms, Allen \& McCray 2000). 
Except where indicated, all quoted errors on spectral parameters are extrema on the two-interesting-parameter, 68\%-confidence contours.

We were able to obtain informative spectral fits (\S \ref{s:sp_s3-5}) for the brightest x-ray source in the field and less informative spectral fits (\S \ref{s:sp_s3-10}) for the next brightest.
For the remaining (fainter) X-ray sources, we could only determine three-band X-ray colors (\S \ref{s:xray_cc}). 

\subsubsection{X-ray Spectrum of S3-5 (``Leon X-1'')\label{s:sp_s3-5}}

Only the brightest source---S3-5 (``Leon X-1'')---had sufficient counts to warrant a serious spectral analysis. 
Because this source was about $5\arcmin$ off-axis, its image was sufficiently blurred that we safely ignored effects of pile-up ($<$5\%).
These data, obtained shortly after opening the telescope's forward cover, preceded any damage to the front-illuminated CCDs from low-energy protons scattered from the X-ray mirrors onto the ACIS focal plane during radiation-belt passage. 
(Since this radiation problem was understood, ACIS has always been stowed---out of focal position---during radiation-belt passages.)
Furthermore, accumulated molecular contamination on the ACIS optical-blocking filters was still negligible. 

\begin{figure}
\begin{center}
\includegraphics[scale=.7,angle=-90]{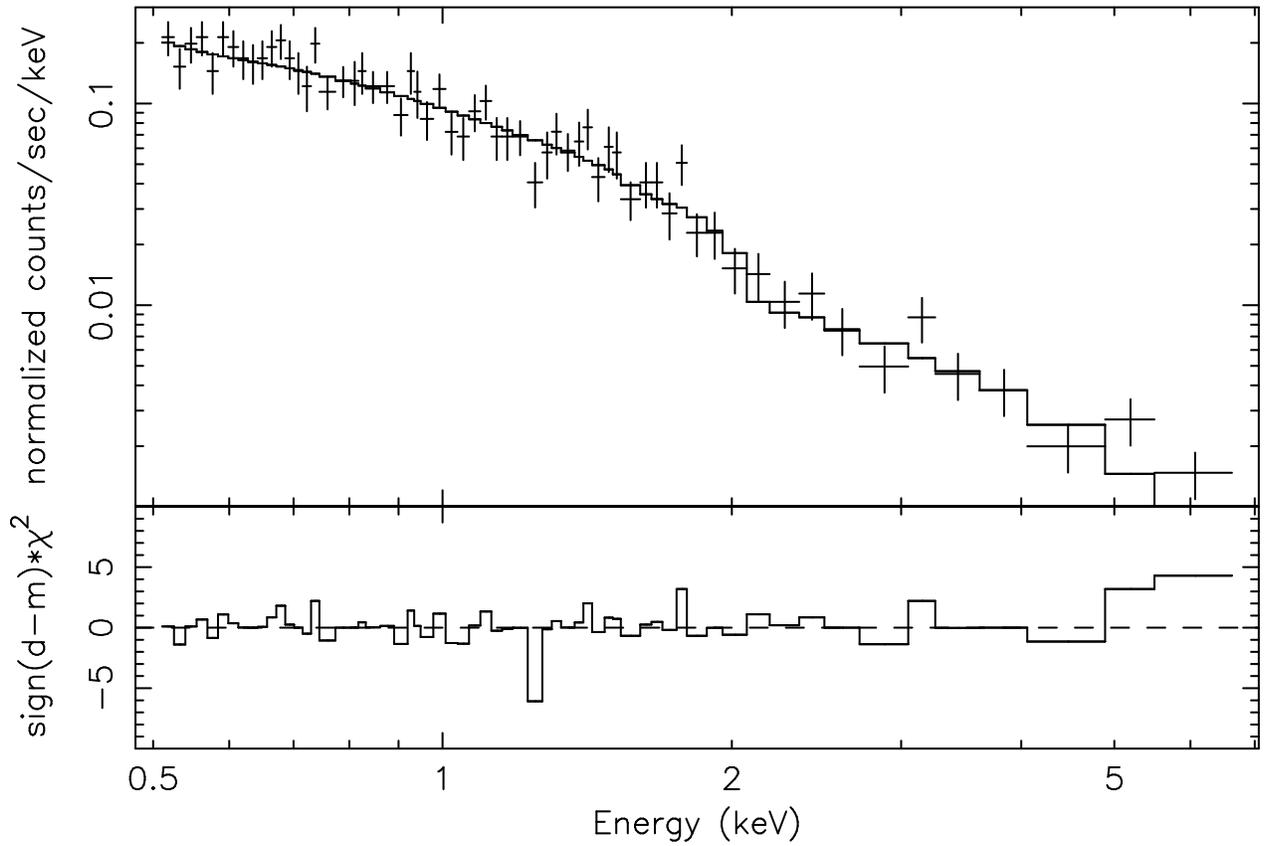}
\end{center}
\caption{Count spectrum of ``Leon X-1'' and fitted model.
The upper panel shows the data with statistical errors and the best-fit power-law model---photon index $\Gamma=2.05$ and negligible absorption.
The lower panel diplays the signed contribution to \chisq\ --- data (d) minus model (m)--- for each energy bin.
\label{f:lx1_spectrum}}
\end{figure}

\begin{figure}
\vskip 0.2in
\begin{center}
\includegraphics[scale=.7,angle=-90]{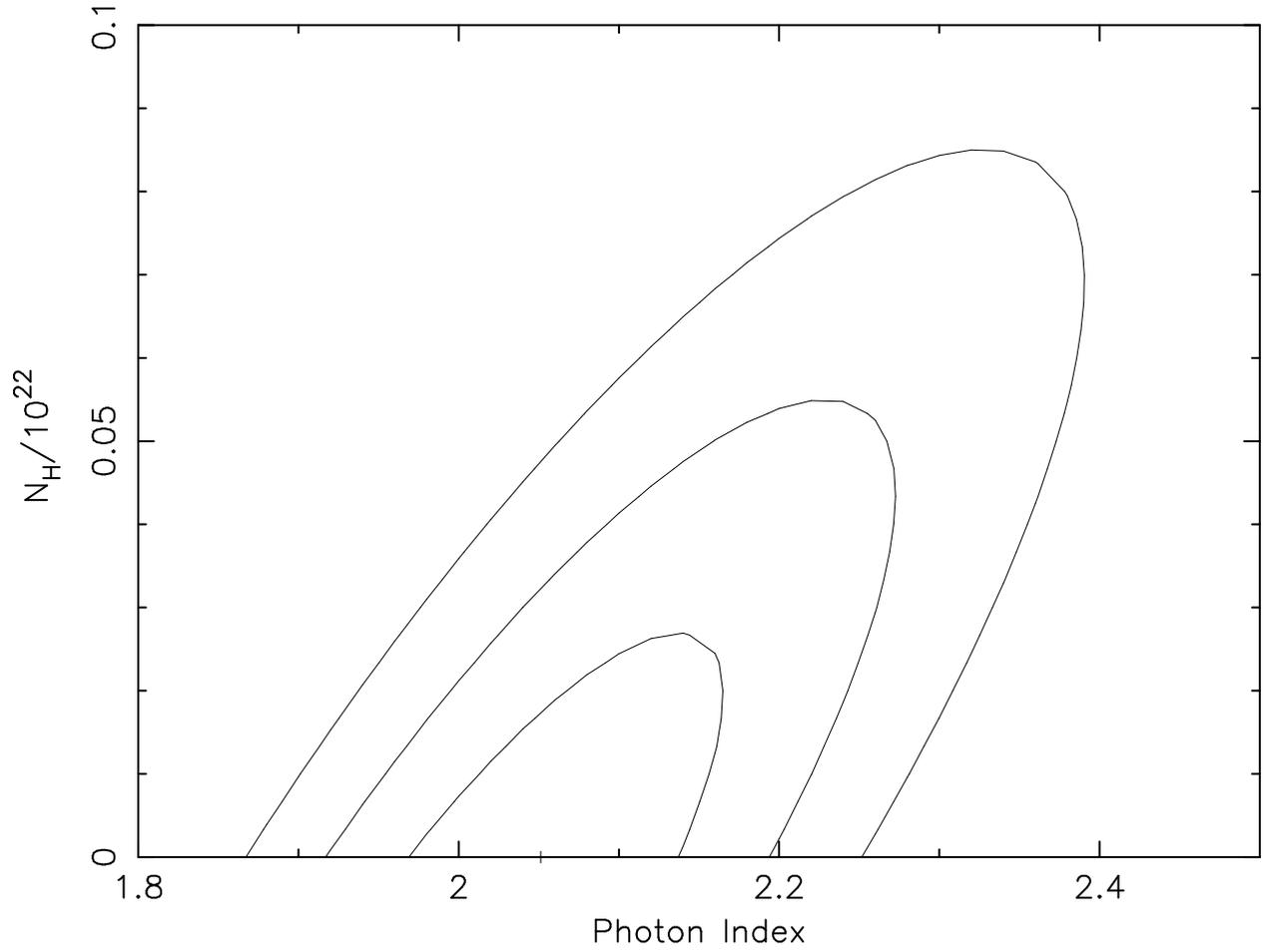}
\end{center}
\caption{Confidence contours for fitting a power law of photon index $\Gamma$ absorbed through a column $N_H$ to the X-ray spectrum of ``Leon X-1''. 
Contours in the $\Gamma\!-\!N_H$ plane denote $\Delta\chi^2=\{2.30, 6.17, 11.8\}$, corresponding to $\{1, 2, 3\}\,\sigma$ significance for two interesting parameters.
\label{f:lx1_contours}}
\end{figure}

In fitting the data, we investigated three XSPEC spectral models---{\tt powerlaw}, {\tt mekal}, and {\tt bbody} (blackbody)---with interstellar absorption {\tt tbabs}. 
Table~\ref{t:spectra} summarizes the results for these fits.
Of the three models, only the power-law model gave a statistically acceptable fit.
This fit requires negligible absorption, consistent with the value ($4\times 10^{20}\, {\rm cm}^{-2}$) inferred from radio data using {\tt colden.}\footnote{http://heasarc.gsfc.nasa.gov/cgi-bin/Tools/w3nh/w3nh.pl}
Figure~\ref{f:lx1_spectrum} displays the best-fit power-law spectrum and the fitting residuals; Figure~\ref{f:lx1_contours}, $\chi^2$-contour plots in the $N_{H}$--$\Gamma$ plane.
Note that the best-fit spectrum---a power-law with photon index $\Gamma \approx 2$---is consistent with a pulsar wind nebula or with an AGN.
\clearpage

\clearpage
Although neither the blackbody model nor the {\tt mekal} model with solar abundances gave an acceptable fit, a two-temperature {\tt mekal} model did:
$\chi^2=54$ on 57 degrees of freedom, with best-fit temperatures $0.25 (+0.04, -0.04)\kev$ and $4.6 (+1.9, -1.1)\kev$.
The higher temperature component accounted for 82\% of the flux, accounting for the absence of lines and  similarity to a thermal-bremsstrahlung model. 
As with the power-law fit, the two-temperature {\tt mekal} fit requires negligible interstellar absorption---$N_H /(10^{22}\, {\rm cm}^{-2}) = 0.0~(+0.1, -0.0)$, for four interesting parameters. 

\subsubsection{X-ray Spectrum of S3-10\label{s:sp_s3-10}}

The next brightest source---S3-10---had only about 160 counts, which we grouped into spectral bins of at least 15 counts each. 
With $\chi^2 = 10$ on 4 degrees of freedom, a power-law fit was marginally unacceptable at 95\% confidence. 
Blackbody and {\tt mekal} models each yielded an acceptable fit, but with very different values for all parameters (Table~\ref{t:spectra}) except the flux.

\subsubsection{X-ray Color--Color Relation\label{s:xray_cc}}

\begin{figure}
\begin{center}
\includegraphics[scale=.7,angle=-90]{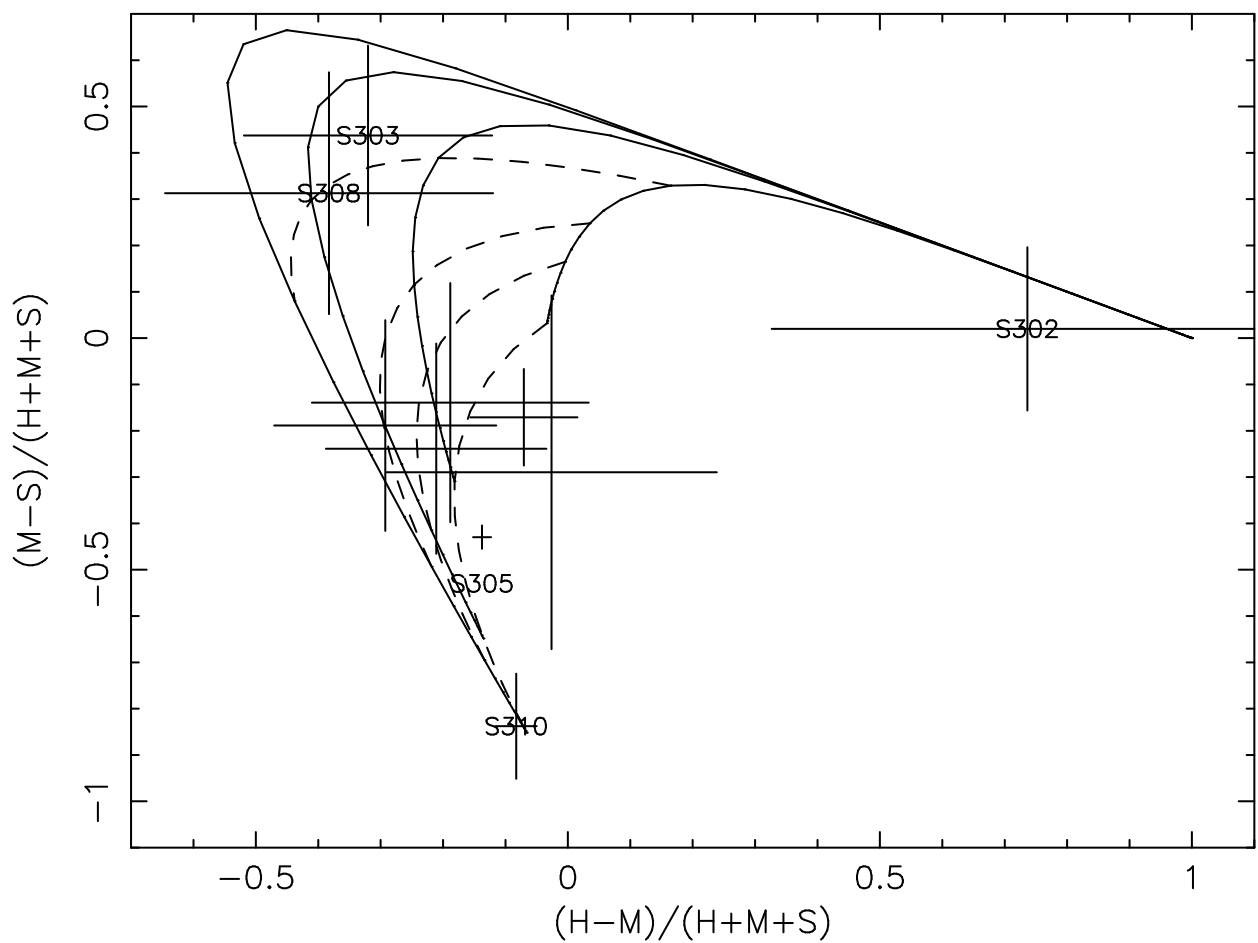}
\end{center}
\caption{X-ray color--color diagram for sources detected with 
(back-illuminated) CCD S3. 
The energy bands are S ($0.5\!-\!1 \kev$), M ($1\!-\!2\,\kev$), and H ($2\!-\!8\,\kev$). 
Solid lines represent colors of absorbed power-law spectra for photon index $\Gamma=\{1, 2, 3,4 \}$ (from innermost to outermost) and varying column $N_H$.
Dashed lines indicate these colors for $N_H=\{0.1,1,2,5\}\!\times\!10^{21}\, {\rm cm}^{-2}$ (innermost to outermost) and varying $\Gamma$. 
\label{f:S3_xcolors}}
\end{figure}

The remaining 13 \chandra\ sources had fewer than 100 detected source counts each; thus, we attempted no spectral fitting for them individually.
However, to estimate the flux (Table~\ref{t:xsources}) of each (faint) source on S3 (1--4 and 6--9), we fit their co-added spectra to an absorbed power-law. This fit yielded a spectral index of 1.45 and a columns of $N_H /(10^{22}\, {\rm cm}^{-2}) = 0.07$. We then scaled the flux of each source proportionally to its detected counts.
For each S3 source, we also determined count rates in three X-ray bands, which Figure~\ref{f:S3_xcolors} displays in an X-ray color--color diagram.
\clearpage
\clearpage
\subsection{Summary of X-ray Properties\label{s:xray_desc}}

To conclude the discussion of the X-ray observation of the first \chandra\ field, we summarize the X-ray properties of the detected sources.

\subsubsection{S3-5 (CXOU J051936.3-604804)\label{s:S3-5}}

With a (0.5--8\,$\kev$) flux of $6.8\!\times\!10^{-13}$ \ergcms, source S3-5 (``Leon X-1'') was by far the brightest X-ray source in the first \chandra\ field. 
For a sky density of about 2 such sources per square degree (Kim et al.\ 2004), it was somewhat fortuitous that so bright a source lay in the initial field of view (0.12 deg$^2$) and on CCD S3 (0.02 deg$^2$). 
We have identified this X-ray source with a 17$^{\rm th}$ B-magnitude Type-1 AGN at redshift $z=0.3207$ (\S \ref{s:vis_sp_res}). 
Based upon its X-ray spectrum (\S \ref{s:sp_s3-5}, Figure~\ref{f:lx1_spectrum}) and redshift (\S \ref{s:vis_sp_res}), the 0.5--8-$\!\kev$ (rest-frame) luminosity of ``Leon X-1'' is about $2\times 10^{44}$ \ergl\ ($H_{o}=70\ {\rm km}\,{\rm s}^{-1}\,{\rm Mpc}^{-1}$, $\Omega_{m}=0.3$, $\Omega_{\Lambda}=0.7$). 

We used \chandra\ spectral data and PIMMS\footnote{http://heasarc.gsfc.nasa.gov/Tools/w3pimms.html} to estimate the flux of ``Leon X-1'' in the \ROSAT\ band.
During the \chandra\ exposure, the inferred \ROSAT\ flux was consistent with that of 1RXS J051934.0-604800 (\S \ref{s:other_x-ray}) in the \ROSAT\ All-Sky Survey Bright-Source Catalog.
Thus, the \chandra\ and \ROSAT\ sources are almost certainly the same object.    

\subsubsection{S3-10 (CXOU J052028.2-604648)\label{s:S3-10}} 

The counterpart to S3-10 appears to be a $7^{\rm th}$-magnitude (A3-V) early main sequence star (\S \ref{s:counterparts}).
However, it is unlikely that an early A star is the source of X rays (e.g., Simon, Drake \& Kim 1995; Schmitt 1997; Daniel, Linsky \& Gagne 2002):
Such stars are too hot to drive convectively a corona (as for G, K, M stars) but too cool to drive a stellar wind (as for O and early B stars).
Thus, if the bright A star is not obscuring an unrelated object, the observed X-ray emission probably comes from a binary companion to the A star. 
In this scenario, the soft X-ray spectrum suggests that the system is not a close binary, which would effect a high stellar rotational frequency and hard-X-ray emission. 

\subsubsection{S3-9 (CXOU J052003.7-604316)} 

The near-infrared magnitudes and colors (Figure~\ref{f:2MASS_colors} and Table~\ref{t:2MASS_colors}) of the 2MASS candidate counterpart to source S3-9 are quite similar to those of source S3-5 (\S \ref{s:S3-5}), which we identified as a Type-1 AGN (\S \ref{s:vis_sp_res}).  
However, the X-ray flux (Table~\ref{t:xsources}) from S3-9 is only 0.02 that from S3-5. 

\subsubsection{S3-3 (CXOU J051912.0-604359) \& S3-8 (CXOU J051959.9-604759)} 

Sources S3-3 and S3-8 have similar X-ray colors (Figure~\ref{f:S3_xcolors}), distinctly different from those of the other X-ray sources in the first \chandra\ field.
These colors are indicative of a moderately large absorption column. 
Furthermore, neither source has an USNO or 2MASS candidate counterpart. 
Thus, they may be obscured AGNs.

\subsubsection{S3-2 (CXOU J051907.1-604500 \& S3-2)}

Source S3-2 is the second faintest X-ray source detected on CCD S3; thus its X-ray colors have rather large errors.
If the X-ray colors (Figure~\ref{f:S3_xcolors}) are correct, its spectrum is much harder than any other source on S3, possibly indicating a very high absorption column.
This and the absence of either a USNO or 2MASS candidate counterpart are consistent with a highly obscured AGN.

\subsubsection{Other X-ray Sources} 

The remaining sources detected on CCD S3 have X-ray colors (Figure~\ref{f:S3_xcolors}) similar to those of S3-5 and S3-9.
Thus, it is tempting to argue that these X-ray sources are also (unobscured) Type-1 AGNs. 
Based upon the large number of AGNs observable with \chandra\ (e.g., Kim et al.\ 2004), this conclusion is reasonable.  

\section{Visible-Light Spectroscopy of ``Leon X-1''\label{s:vis_sp}}

Using spectrographs of the European Southern Observatory (ESO) at La Silla (Chile), we obtained visible-light spectroscopy of the candidate counterpart to the field's brightest X-ray source (S3-5, ``Leon X-1'').  
We briefly describe (\S \ref{s:vis_sp_obs}) those spectroscopic observations and then discuss (\S \ref{s:vis_sp_res}) the results.

\subsection{Spectroscopic Observations\label{s:vis_sp_obs}}

For the spectroscopy, we used both the ESO Multi-Mode Instrument imaging spectrograph (EMMI, \S \ref{s:vis_emmi}) and the ESO Faint Object Spectrograph and Camera (EFOSC2, \S \ref{s:vis_efosc2}).
Due to different capabilities of the two instruments, the two observations were complementary (\S \ref{s:vis_comp}).

\subsubsection{EMMI Spectrum\label{s:vis_emmi}}

\begin{figure}
  \begin{center}
\includegraphics[angle=90,width=16cm]{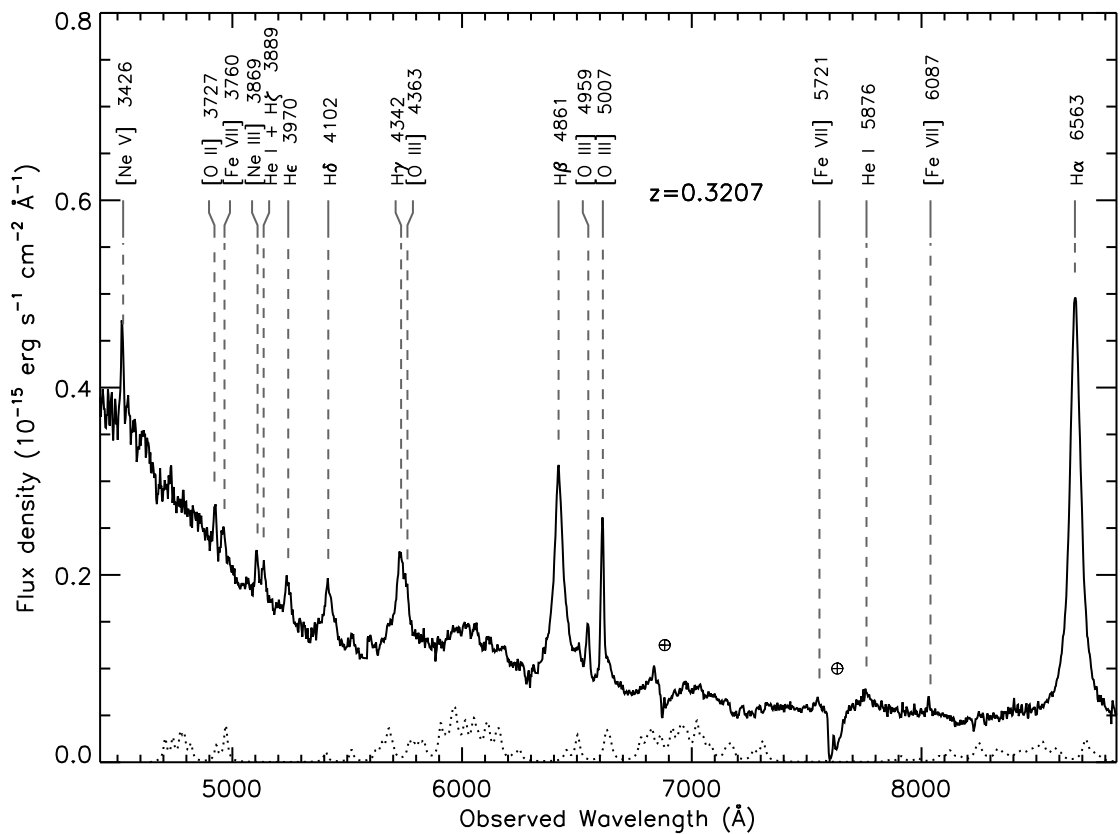}
\caption{EMMI flux-calibrated spectrum of ``Leon X-1'', with identification of the strongest emission lines. 
Circled crosses mark the strongest telluric (terrestrial) absorption features. 
The dotted curve shows a pure-\feii\ template spectrum (V\'{e}ron-Cetty, Joly \& V\'{e}ron 2004), redshifted ($z=0.3207$) and sampled at the dispersion of the EMMI spectrum.} \label{f:emmispec} 
  \end{center}
\end{figure}

We used the EMMI to obtain a spectrum of ``Leon X-1'' (Figure~~\ref{f:emmispec}), on the morning of 2004 March 23. 
The EMMI grism used for this observation spans 3100--9000~\AA\ with a 2.86-\AA/pix dispersion and 8-\AA\ (FWHM, Full-Width-at-Half-Maximum) resolution. 
Initially, the primary purpose of the EMMI spectrum was to determine the redshift. Consequently, we utilized available telescope time to obtain a 1800-s exposure at an air mass greater than four (4).
In order to minimize effects of differential refraction at large air mass, we obtained the spectrum at parallactic angle---i.e., with dispersion direction parallel to horizon.
Furthermore, we  corrected the spectrum for average atmospheric extinction using  Cerro Tololo Inter-American Observatory (CTIO) coefficients and calibrated the flux using two reference stars---albeit, at relatively low air mass (1.03 and 1.18). 
We estimate that chromatic errors in the line-flux measurements are less than 5--10\%\ longward of 4500 \AA.
\clearpage
\clearpage
\subsubsection{EFOSC2 Spectrum\label{s:vis_efosc2}}

\begin{figure}
  \begin{center}
\includegraphics[angle=90,width=16cm]{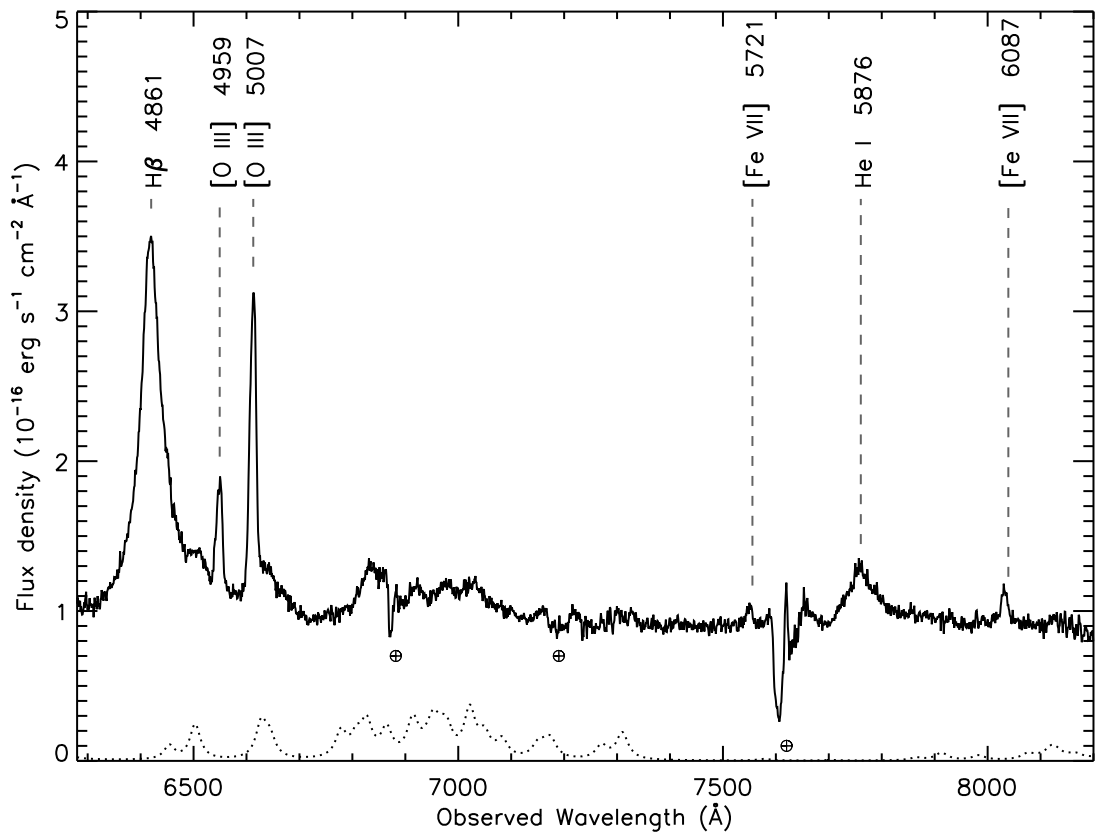}
\caption{EFOSC2 flux-calibrated spectrum of ``Leon X-1'', with identification of the strongest emission lines. 
Circled crosses mark the strongest telluric (terrestrial) absorption features. 
The dotted curve shows a pure-\feii\ template spectrum (V\'{e}ron-Cetty, Joly \& V\'{e}ron 2004), redshifted ($z=0.3207$) and sampled at the dispersion of the EFOSC2 spectrum.} \label{f:efoscspec} 
  \end{center}
\end{figure}

In order to achieve somewhat better resolution and signal-to-noise ratio, we also observed ``Leon X-1'' with the EFOSC2 at the 3.6-m telescope, on the night of 2005 January 20. 
Under clear conditions and $<\!1.5\arcsec$ seeing, we obtained three (3) 2200-s exposures at a median air mass of 1.25. 
These spectra span 6300--8200 \AA\ at 5.2-\AA\ (5.6-pix) FWHM resolution, with a 20--30 signal-to-noise ratio in the continuum. 
After flat-fielding, 6\% fringing in the red part of the spectrum remained; however, this does not significantly affect any emission line. 
We flux-calibrated the combined spectrum (Figure~\ref{f:efoscspec}) against a standard star (HD60753, spectral type B2III) reduced identically. 
This calibration also corrects for flux lost through the 1\arcsec-wide entrance slit.
\clearpage
\clearpage
\subsubsection{Comparison of EMMI and EFOSC2 Spectra\label{s:vis_comp}}

The EMMI and EFOSC2 spectra are complementary in that the former spans a broader spectral range, whereas the later is of higher quality.
There are some cross-calibration uncertainties between the two spectra, obtained with different instruments under differing conditions. 
Thus, we scaled the EMMI spectrum upwards so that the flux of the narrow \hb\ component matched that in the higher quality EFOSC2 spectrum. 
However, this flux scale factor was only 1.1 and there were no significant differences in the wavelength calibrations.

\subsection{Spectroscopic Results\label{s:vis_sp_res}}

The extensive spectral coverage of the EMMI spectrum includes numerous emission lines---forbidden lines and narrow and broad permitted lines---as identified in Figure~\ref{f:emmispec}. 
Based upon observed wavelengths of the \ha\ and \oiii\,\l\l4959,5007 lines, we obtain a redshift $z=0.3207\pm 0.0004$. 
Furthermore, the blue continuum and broad permitted lines indicate little line-of-sight obscuration to the broad-line region. 
Consequently, we classify ``Leon X-1'' as a Type-1 AGN, consistent with its power-law X-ray spectrum exhibiting little intrinsic absorption. 
Within its wavelength range, the EFOSC2 spectrum (Figure~~\ref{f:efoscspec}) shows the same features as the (wider-range) EMMI spectrum, with somewhat better spectral resolution and higher signal-to-noise.

In the remainder of this section, we describe our analysis of the EMMI and EFOSC2 spectra of ``Leon X-1''---fits to the observed emission lines (\S \ref{s:vis_fits}), the \feii\ line complex (\S \ref{s:vis_feii}), a relative blueshift of higher ionization forbidden lines (\S \ref{s:vis_shift}), and the hydrogen Balmer decrement (\S \ref{s:vis_ratio}).
The following section (\S \ref{s:agn_prop}) then investigates these results in terms of the multi-wavelength properties of AGNs.

\subsubsection{Emission-Line Fits\label{s:vis_fits}}

We fit recognized emission lines with gaussian profiles, after estimating the continuum under each line or each set of simultaneously fitted lines. 
Each of the stronger permitted lines clearly exhibits a narrow core and extended wings, which we fit with double (broad and narrow) gaussian components. 
The spectra also show several emission-line blends---e.g, \feii\ \l4924 with \hb\ (\l4861), \feii\ \l5018 with \oiii\ \l\l4959,5007, and \oiii\ \l4363 with \hg\ (\l4342). 
Table~\ref{t:emmi-lines} lists the fitted emission-line parameters---rest and observed wavelengths, redshift, FWHM, line flux, and flux ratio to \hb$_{N}$ (narrow) and to \hb$_{B}$ (broad) components---for the EMMI spectrum; Table~\ref{t:efosc2-lines}, for the EFOSC2 spectrum.
Where possible, we used the (higher quality) EFOSC2 data to model better blended lines (Table~\ref{t:efosc2-lines}).

The FWHMs of the two strongest broad lines---\ha\ and \hb---are about $4000\ {\rm km\, s^{-1}}$ (Table~\ref{t:emmi-lines}).
Those of the other (weaker) broad lines scatter about this value (Tables~\ref{t:emmi-lines} and \ref{t:efosc2-lines}), being generally less accurate due to lower signal and to blending.
The FWHMs of the narrow Balmer components are $1000\!-\!1500\ {\rm km\, s^{-1}}$; those of the remaining lines---all forbidden lines and the narrow component of \hei\ \l5876---are smaller. 
Except for a possible blue extension in each line of the \oiii\ \l\l4959,5007 doublet, no line departs significantly from gaussian, within limits imposed by line blending and signal-to-noise.

\subsubsection{\feii\ Line Complex\label{s:vis_feii}}

The complexes of permitted \feii\ lines---thought to originate through resonance fluorescence and collisional excitation (e.g., Osterbrock 1988)---are among the strongest components of the visible-light emission-line spectra of unobscured AGNs (Wills, Netzer \& Wills 1985).
The EMMI spectrum of ``Leon X-1'' (Figure~~\ref{f:emmispec}) displays at least two strong \feii\ emission blends---near 6000\,\AA\ [4500\,\AA] and near 7000\,\AA\ [5300\,\AA] in the observed [rest] frame, the second being affected by strong telluric absorption near 6800\,\AA. 
The first blend comprises lines mainly from \feii\ multiplets 37 and 38; the   second, from multiplets 48 and 49 (e.g., Phillips 1976; references therein). 
For comparison, Figures~\ref{f:emmispec} and \ref{f:efoscspec} show the (arbitrarily scaled) template \feii\ spectrum of I~Zw~1 (compiled by V\'{e}ron-Cetty, Joly \& V\'{e}ron 2004), redshifted and re-binned as appropriate for ``Leon X-1'' with no additional broadening. 
Of importance for fitting lines, \feii\ \l4924 and \feii\ \l5018 (from multiplet 42) blend with \hb\ and the \oiii\,\l\l4959,5007 doublet. 
Noteably, the flux of the \feii\ complex around 4500 \AA\ (rest) is strong, as discussed below (\S \ref{s:agn_prop}).

\subsubsection{Forbidden-Line Shifts\label{s:vis_shift}}

While its strongest forbidden lines are the \oiii\,\l\l4959,5007 doublet, the spectrum of ``Leon X-1'' exhibits several high-ionization forbidden lines---\fevii\ and \nev---characteristic of a hot or a photoionized environment, such as that of an AGN. 
Typical AGN lines that are not detected in ``Leon X-1'' include \oi\,\l6300 and \sii\,\l4067. 
There is slight evidence for weak \nii\,\l6583, but the fit is not well constrained because of its superposition over the broad and bright wing of \ha.

\begin{figure}
  \begin{center}
\includegraphics[angle=90,width=15cm]{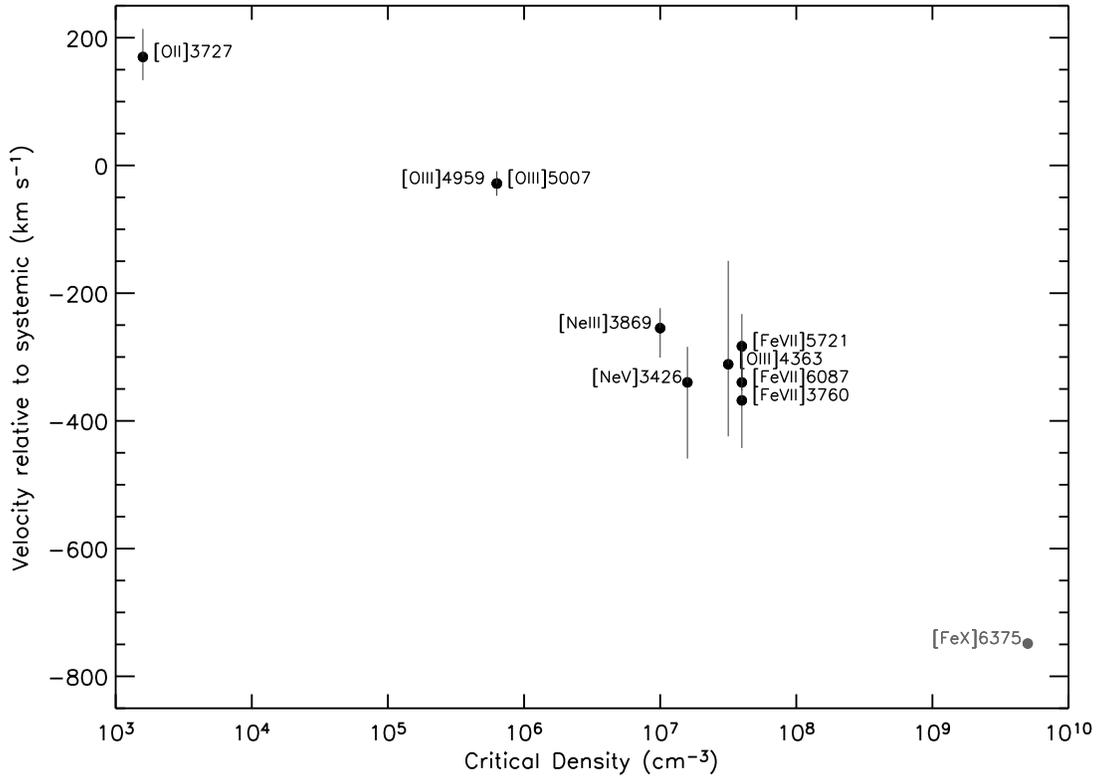}
\caption{Plot of relative redshift of each forbidden line against its critical density for collisional de-excitation.
Plotted redshifts are relative to a systemic redshift of $z=0.3207$, with 1-$\sigma$ errors displayed. 
Critical densities are those for $T=10^4$ K (Appenzeller \& \"{O}streicher 1988). 
The Grey symbol for the \fex\ line indicates a marginal detection.} \label{f:lineshifts} 
  \end{center}
\end{figure}

Interestingly, the higher-ionization forbidden lines are blueshifted relative to the permitted lines, as well as to the low-ionization forbidden lines. 
Figure~\ref{f:lineshifts} shows a correlation of relative blueshift of each detected forbidden line, with its critical density for collisional de-excitation. 
In that no stellar or interstellar absorption features are evident, we assume a systemic redshift of $z=0.3207$ (\S \ref{s:vis_sp_res}). 
\clearpage
\clearpage
Other authors (e.g.,  Pelat, Alloin, \& Fosbury 1981) have noted such a correlation for other AGNs, inferring that the higher-ionization forbidden lines probably originate in photoionized clouds between the broad-line and the narrow-line regions.
Thus, their relative proximity to the nucleus would result in higher velocities and ionization parameters relative to the those of the narrow-line regions.
However, such Seyfert galaxies usually display significant line-profile asymmetries toward the blue, implying both outflow and extinction of emission from the far hemisphere of the AGN. 
In contrast, most line profiles for Leon X-1 are symmetric, with only \oiii\,\l\l4959,5007 showing a weak asymmetry (\S \ref{s:vis_fits}). 
This argues against a spherically-symmetric outflow (e.g., Appenzeller \& \"{O}streicher 1988), but would be consistent with bipolar ejection, if radiation from receding material is suppressed---e.g., extinguished--- while that from approaching material is not. 

\subsubsection{Hydrogen Balmer Decrement\label{s:vis_ratio}} 

The Leon X-1 spectrum shows a prominent hydrogen Balmer sequence, from \ha\ through \he.  
Generally interpreted as recombination emission lines from photoionized gases near the AGN's core, their relative strengths provide diagnostics of physical conditions in these regions. 
Our analysis of the Leon X-1 spectrum finds a Balmer decrement \{\ha, \hg, \hd\}/\hb\ of \{1.56, 0.41, 0.33\} and \{3.79, 0.40, 0.18\}, respectively, for the broad-line and the narrow-line series (Table~\ref{t:emmi-lines}) .  
Photoionization--recombination models (e.g., Osterbrock 1989; references therein) typically give \{\ha, \hg, \hd\}/\hb\ $\approx$ \{3, 0.5, 0.3\}, for the (unreddened) decrement. 
For Leon X-1, the observed decrement for the narrow-line Balmer sequence is consistent with this case and a small amount of reddening (cf.\ Figure~1 in Osterbrock, Capriotti \& Bautz 1963). 
   
Unlike the narrow-line Balmer decrement, the broad-line decrement is inconsistent with standard photoionization--recombination cases (A and B; Osterbrock 1989), for any value of the reddening. 
For the broad-line component in Leon X-1, the measured \ha /\hb\ $\approx 1.6$.
Systematic uncertainties in our observations---e.g., due to line blending and calibration errors---might raise this ratio to about 2.0, still significantly below the canonical value for a tenuous photoionized plasma. 
Furthermore, this ratio (\ha /\hb\ $\approx 1.6$) is unusually small compared to that found in other AGNs (e.g., Vaughan et al.\ 2001).
Reddening, of course, can only steepen the decrement and thus cannot account for so low an \ha /\hb\ ratio.
Other effects---self-absorption in the Balmer lines, collisional excitation and de-excitation, stimulated emission, etc.---can also alter the Balmer decrement (Osterbrock, Capriotti \& Bautz 1963; Capriotti 1964; Cox \& Mathews 1969; Gerola, Salem \& Panagia 1971; Parker 1964; Adams \& Petrosian 1974; Netzer 1975, 1977;  Krolik \& McKee 1978; Ferland \& Rees 1988; Rees, Netzer \& Ferland; Zheng \& Puetter 1990).
These studies (especially Krolik \& McKee 1978; Drake \& Ulrich 1980; Zheng \& Puetter 1990) indicate that the most favorable conditions for producing a flat Balmer decrement are non-negligible optical depths in \ha\ and high densities, which increase collisional de-excitation, driving the hydrogen-level population toward local thermodynamic equilibrium.
In the extreme, these conditions produce a very flat Balmer decrement, as observed in the spectra of cataclysmic-variable accretion disks (Williams 1983; Elitzur et al.\ 1983; Williams \& Shipman 1988)

Consequently, the broad-line Balmer decrement observed in Leon X-1 suggests emission from dense photoionized gas in which collisional de-excitation is non-negligible.
While the emitting gas could be in cloudlets circulating in the inner region of the AGN, it could also comprise the photosurface of an accretion disk (see also Collin-Souffrin et al.\ 1981; Collin-Souffrin, Dumont \& Tully 1982).  
If we accept the large black-hole mass and high accretion rate inferred from our analysis below (\S \ref{s:agn_prop}), this broad-line emitting gas may reside in the distal region of the accretion disk that fuels the UV/X radiation from black holes.  

\section{Properties of Leon X-1 in the Context of the E1 Correlation Space\label{s:agn_prop}} 

Analyses of multi-wavelength parameters of AGNs show correlations amongst widths and strengths of \hb, \oiii\,\l 5007, and \feii\ optical emission lines, and soft-X-ray photon index. 
A remarkable feature of the ``Eigenvector 1'' (E1) correlation space in a Principal-Component Analysis (Boroson \& Green 1992, hereafter BG92; Wang, Brinkman \& Bergeron 1996; Sulentic et al.\ 2000b) is that radio-loud (RL) and radio-quiet (RQ) AGNs occupy very different regions in the E1 projected planes (Sulentic et al.\ 2003).   
The proposed physical drivers for these correlations are supermassive-black-hole mass, accretion rate, and system orientation.  
Furthermore, a recent study (Zamanov \& Marziani 2002) has shown that these same correlations apply to stellar-mass accreting systems---e.g., interacting binaries in which the accretor is a white dwarf. 
In view of the apparent robustness of the E1 correlation space, we use it in examining the X-ray, visible-light, and radio data for Leon X-1, in order to estimate its black-hole mass and accretion rate. 
Based upon its redshift ($z = 0.32$) and weak radio emission---4.5-mJy SUMSS (Sydney University Molonglo Sky Survey) upper limit (Richard Hunstead, private communication)---Leon X-1 is a RQ AGN.   
 
In the BG92 sample, the broad-line-\hb\ FWHM is typically 11000--20000\, ${\rm km}~{\rm s}^{-1}$ for RL AGNs, but much smaller---mean $\approx 2000\, {\rm km}\,{\rm s}^{-1}$---for RQ AGNs (Sulentic et al.\ 2000). 
Sulentic et al.\ (2000) categorize RQ AGNs with somewhat larger broad-line-\hb\ FWHMs---3000--4000~${\rm km}~{\rm s}^{-1}$---as population ``A''.
For these population-A RQ AGNs, the mean equivalent width [standard deviation] of broad-line \hb\ emission is EW(\hb$_{B}$) $\approx 128$\,\AA\ [40~\AA]; that of \oiii\,\l 5007 is EW(\oiii\,\l 5007) $\approx 20$\,\AA\ [20~\AA].
Sulentic et al.\ (2002) also find the mean [standard deviation] absolute B magnitude of the population-A RQ AGNs in the BG92 sample to be $M_B = -22.3$ [1.8].
We find that the corresponding parameters for Leon X-1---FWHM(\hb$_{B}) \approx 3757\, {\rm km}\, {\rm s}^{-1}$, EW(\hb$_{B}$) $\approx 118$\,\AA, EW(\oiii\,\l 5007) $\approx 26$\,\AA, and $M_B \approx -24$---are characteristic of a population-A RQ AGN.

On the other hand, the other two E1 parameters---namely, the soft-X-ray photon index $\Gamma$ and the strength of the \feii-complex at 4570~\AA---reveal that Leon X-1 is not a typical Population-A RQ AGN. 
In particular, we find $\Gamma = 2.05^{+0.11}_{-0.08}$, versus the BG92-sample population-A mean [standard deviation] of $\Gamma = $ 2.6 [0.1] (Sulentic et al.\ 2000a).
%
%
However, the steep mean photon index found for the BG92-sample is based on ROSAT
data (0.3 - 2.4 keV) and largely reflects the soft excess which is common in
low redshift ($z < 0.4$) quasars Porquet et al. (2004). The mean hard photon
index (2 - 12 keV) for a sample of 40 PG quasars based on XMM spectra is
$1.89\pm0.11$ Piconcelli et al. (2005), fully consistent with our measured value.
Leon X-1 is therefore somewhat unusual in the absence of a soft
excess.\footnote{This observation was taken with a focal plane CCD temperature
of -90\,C, prior to any measurements of the internal calibration source.  The
CCD gain in this configuration is therefore not well-calibrated.  However, even
a substantial error in the gain would not affect the power law index or the
lack of an observed soft excess. We also analyzed the spectral data using both the responses based on calibration data base 3.1.0 using CIAO 3.2.2 for -90 and for -120 \,C. The best fit parameters were essentially identical with differences well below the statistical errors.}
In addition, we find EW(\feii\,\l 4570) = 160~\AA\ for Leon X-1, versus EW(\feii\,\l 4570) = 60~\AA\ [16~\AA].
AGNs with EW(\feii\,\l 4570) larger than 100~\AA\ are rare (Lipari, Terlevich \& Macchetto 1993; Grupe et al.\ 1999). 
Ultra-strong Fe emitters are typically broad-absorption-line (BAL) QSOs or strong infrared Seyfert-1 galaxies (e.g., Hartig \& Baldwin 1986; Zheng et al.\ 2002; Yuan \& Wills 2003).  
Such strong \feii\ emission may result from either a high accretion rate onto a very massive central accretor or from active star formation. 

Comparison of Leon X-1 with the BG92 AGN sample in the principal correlation planes of the E1 space (Figure~1 in Sulentic et al.\ 2000b) confirms that it is distinguishable from other AGNs primarily because of its strong \feii\ emission.  
Leon X-1 is among the population-A RQ AGNs with the largest EW(\feii) and EW(\feii)/EW(\hb$_{B}$)---population A3/A4 in the optical-parameter plane of the E1 space (see Sulentic et al.\ 2002). 
However, the strength of the \oiii\ lines relative to the \hb\ broad component is more similar to population-A1/A2 RQ AGNs and narrow-line Seyfert-1 galaxies (cf.\ Figure~\ref{f:emmispec} for Leon X-1 with Figure~2 in Sulentic et al.\ 2002).
\clearpage
\begin{figure}
\vskip 0.2in
\begin{center}
\includegraphics[angle=-90]{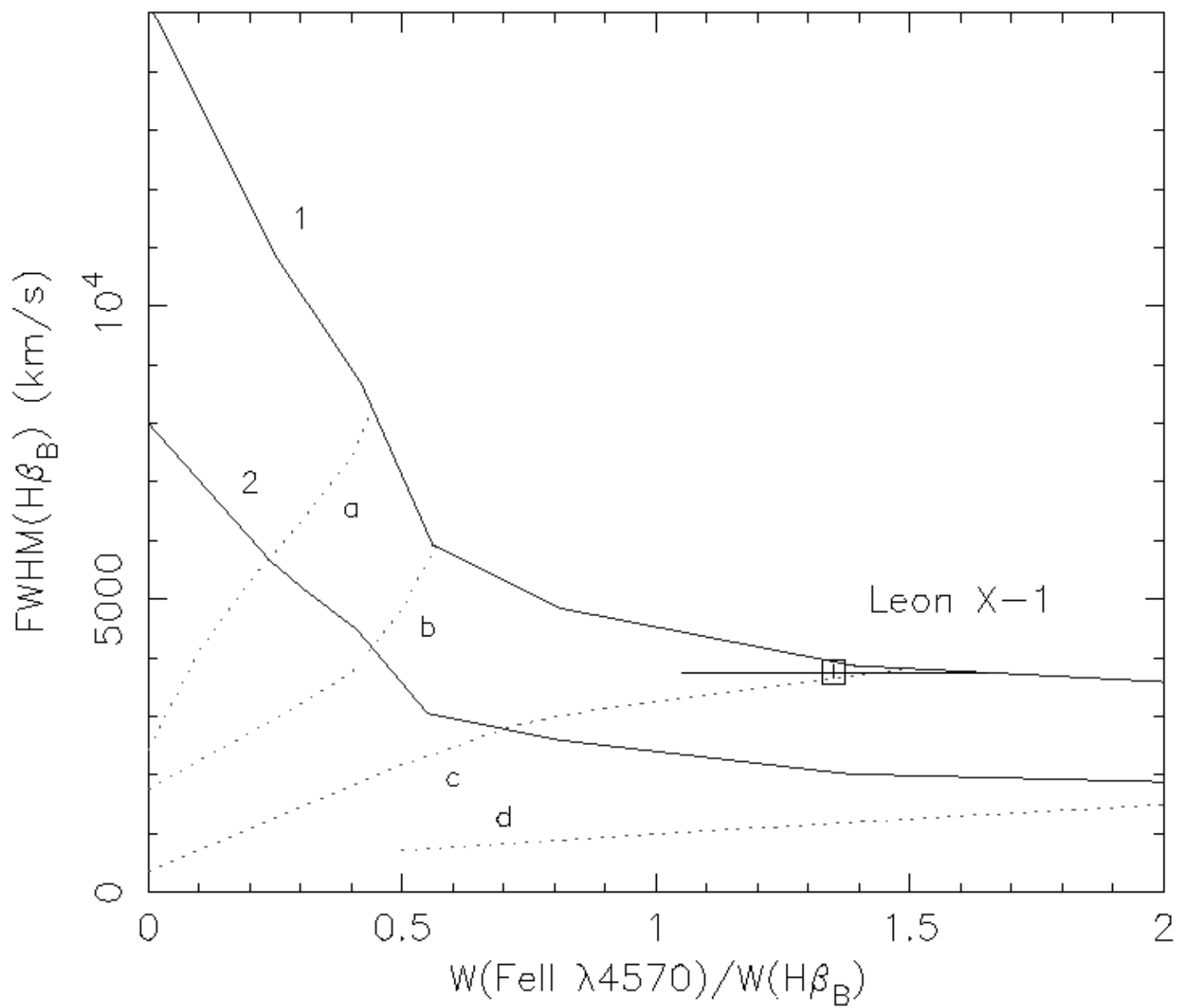}
\end{center}
\caption{Projection of the Eigenvector-1 (E1) correlation space onto the EW(\feii\, \l 4570)/EW(\hb$_{B}$)--FWHM(\hb$_{B}$) plane. 
Solid lines \{1, 2\} are theoretical contours for systems with black-hole masses $\log (M/M_\odot) =$ \{9.5, 8.5\}. 
Dotted lines \{a, b, c, d\} correspond to effective accretion rates $\log [(L/M)/(L_\odot/M_\odot)]=$ \{3.3, 3.7, 4.2, 4.8\}.   
The model grid is adapted from Zamanov \& Marziani (2002).
\label{f:e1-plane}}
\end{figure}
\clearpage 
Figure~\ref{f:e1-plane} plots Leon X-1 in the EW(\feii\, \l 4570)/EW(\hb$_{B}$)--FWHM(\hb$_{B}$) plane, along with a model grid from Zamanov \& Marziani (2002).
The grid coordinates correspond to values (Figure-\ref{f:e1-plane} caption) of the accretor mass and the mass-scaled (bolometric) luminosity---i.e., effective accretion rate---for a model of accretion onto a black hole.
Applying this parameterized model to Leon X-1, we would infer a black-hole mass exceeding $10^{9}\, M_{\odot}$ and an effective accretion rate $10^{4}\, L_{\odot}/M_{\odot}$---only about a factor of three (3) below the Eddington limit. If we take these values as representative, the bolometric luminosity of Leon X-1 would be about $5 \times 10^{46}$ \ergl\ for a $10^{9}\,M_\odot$ accretor---200 times greater than the (0.5--8\,$\kev$) X-ray luminosity we observed. 
Reconciling these luminosities would require that most of the bolometric luminosity from Leon X-1 lies outside the X-ray band.
For a $10^{9}\,M_\odot$ black hole accreting at a rate similar to our estimate for Leon X-1, the peak of the accretion-disk spectrum lies below 30~eV (cf.\ Figure 7.6 in Krolik 1999). 

The multi-wavelength properties of Leon X-1 are unusual compared to more common AGN.
Some of these differences might result from a rare combination of large black-hole mass, a high accretion rate, and perhaps active star formation. 
In the context of E1 correlation space, Leon X-1 is an outlier together with PG 1351+236, PG 0043+038, Mark 231 and 0759+651 (see Fig. 1 of Marziani et al. 2001). 
The four sources were found to show substantial mid and far IR emission and a significant rise toward the far IR, suggesting significant contribution from circumstellar star formation. 
The excess of Fe II in Leon X-1, as in Mark 231 and 0759+651, in comparison with other BL QSO, requires additional excitation processes. 
One possible mechanism is the shocks associated with star formation. 
Leon X-1 has the optical properties of a Type 1 AGN, and this could be the rare situation where the central quasar is "dusty" as well as "naked" as pointed out by Haas et al. (2000). 
We also note that the star formation in such AGN could be caused by jet induced cloud collapse as per Saxton et al. (2005). 
In this case the induction of star formation by quenching of the jets/outflow (usually associated with high L/M) provides an explanation for strong radio emission from the source.  

In summary, Leon X-1 is a rather extreme system amongst RQ AGNs.  
Our multi-wavelength data suggest that it may be a very massive black hole  accreting at a rate close to its Eddington limit. 
However, the absence of significant internal absorption does not evidence a dense outflow, which would be expected for such a system.
Furthermore, the observed X-ray luminosity is less than a percent of the bolometric model implied by the model.  
Alternatively, starburst activity might account for some of its extreme properties.
Broad-band IR/visible/UV observations would help to address these issues.

\acknowledgments

MCW thanks Marshall Joy and Roc Cutri for clarifying information on the infrared spectrum, Gordon Garmire for helping reconstruct the timeline for this observation, and Doug Swartz and Jonathan Grindlay for providing comments and encouragement.
We acknowledge support from the \chandra\ Project and use of the ESO facilities at La Silla (Chile).  
In particular, we thank former La Silla Director Jorge Melnick for use of the EFOSC2.
This publication utilizes some data products from the Two-Micron All Sky Survey,
which is a joint project of the University of Massachusetts and the Infrared
Processing and Analysis Center/California Institute of Technology, funded by
the NASA and by the NSF. 
All of us appreciate the tremendous contributions of Leon Van Speybroeck to the
\chandra\ project and to X-ray astronomy: 
We dedicate this paper to his memory.

\clearpage

\begin{deluxetable}{llrrrrccc}
\tabletypesize{\small}
\tablewidth{0pc}
\tablecaption{X-ray sources in the first \chandra\ field.\label{t:xsources}}
    \tablehead{(1) & (2) & (3) & (4) & (5) & (6) & (7) & (8) & (9)}
   \startdata
Designation      & X-ray & $r_{1}^{a}$& N$^{b}$ & S/N$^c$ & $\sigma_X^d$&           Flux$^e$ &  USNO$^f$ & 2MASS$^f$ \\ 
CXOU             & source &($\arcsec$)&         &         & ($\arcsec$) &  ($F_{15}$)&($\arcsec$)&($\arcsec$)\\ \hline\\[-2.0ex]
J051903.8-604401 & S3-1  &      3.0   &     8.5 &     2.8 &        0.69 &                  6 &       0.9 &           \\      
J051907.1-604500 & S3-2  &      2.9   &    11.0 &    3.0  &        0.61 &                  8 &           &           \\      
J051912.0-604359 & S3-3  &      2.1   &    27.7 &    4.6  &        0.39 &                 20 &           &           \\      
J051917.9-603316 & S2-1  &      1.5   &    32.2 &    4.5  &        1.55 &               $^g$ &       0.4 &      0.40 \\      
J051932.6-604619 & S3-4  &      2.4   &    21.6 &    4.5  &        0.43 &                 15 &           &           \\      
J051936.3-604804 &S3-5$^h$&     4.2   &  1974.0 &     41  &        0.31 &                680 &      0.60 &      0.5  \\      
J051943.2-604142 & S3-6  &       1.5  &    83.2 &    7.6  &        0.32 &                 59 &      0.40 &           \\      
J051958.2-604533 & S3-7    &           2.9     &
22.2    &       4.5     &       0.48    &       16      &               &
\\      
J051959.9-604759 & S3-8    &         5.3    &
17.7    &       3.8     &       0.82    &       13      &               &
\\      
J052003.7-604316 & S3-9    &         2.8     &
16.4    &       3.8     &       0.51    &       12      &       0.40    &
0.3     \\      
J052028.2-604648 & S3-10    &         8.3      &
165     7&      11      &       0.49    &       35      &       0.10    &
0.1     \\      
J052031.5-604002 & IA-1     &         8.7    &
15.3    &       3.2     &       1.37    & $^g$          &       1.50    &
\\      
J052116.9-604430 & IA-2     &        21.0      &
19.7    &       4.3     &       2.94    & $^g$          &               &
\\      
J052116.9-604100 & IA-3     &        22.0      &
64.3    &       6.8     &       1.68    & $^g$          &       2.7\&1.2
&               \\      
J052144.5-602918 & IA-4     &       59.0     &
94.7    &       6.9     &       3.65    & $^g$          &               &
\\ 
\enddata
\\[-4ex] 
\tablecomments{\\
$^a$ Source extraction radius\\
$^b$ Approximate number of source counts (after background subtraction)\\
$^c$ Detection Signal-to-Noise ratio\\
$^d$ X-ray position uncertainty ($1\sigma$ radius, as discussed in the text)\\
$^e$ X-ray (0.5--8.0\,-$\!\kev$) flux in units of $10^{-15}$ \ergcms\\
$^f$ Separation between X-ray position and cataloged position of candidate counterpart\\
$^g$ Discussed in the text\\
$^h$ Brightest source in the first \chandra\ field, dubbed ``Leon X-1''\\}
 \end{deluxetable}

\begin{deluxetable}{lccccccc}
\tabletypesize{\small}
   \tablewidth{0pc}
   \tablecaption{Candidate counterparts to \chandra\ sources. \label{t:counterparts}}
    \tablehead{(1) & (2) & (3) & (4) & (5) & (6) & (7) & (8)}
   \startdata
X-ray & \multicolumn{3}{c}{USNO}   & \multicolumn{3}{c}{2MASS} & Diff.\\
source & RA    & Dec & $N_{r99}^a$ & RA & Dec & $N_{r99}^a$ & $\delta^b$ \\
 & (J2000) & (J2000) &  & (J2000) & (J2000) & & ($\arcsec$)\\ \hline \\[-2.5ex]
S3-1 & 79.766284 & -60.733748 & 0.0089 &  &  & 0.0234 & \\
S3-2 &  &  & 0.0072 &  &  & 0.0189 & \\
S3-3 &  &  & 0.0029 &  &  & 0.0075 & \\
S2-1 & 79.824812 & -60.554545 & 0.0456 &  &  & 0.1197 & \\ 
S3-4 &  &  & 0.0035 &  &  & 0.0093 & \\
S3-5 & 79.901437 & -60.801170 & 0.0018 & 79.901378 & -60.801132  & 0.0047 & 0.17 \\
S3-6 & 79.930200 & -60.695042 & 0.0029 &  &  & 0.0077 & \\
S3-7 &  &  & 0.0044 &  &  & 0.0115 & \\
S3-8 &  &  & 0.0127 &  &  & 0.0333 & \\
S3-9 & 80.015428 & -60.721203 & 0.0050 & 80.015043 & -60.721169 & 0.0131 & 0.69 \\
S3-10 & 80.117503 & -60.779848 & 0.0046 & 80.117507 & -60.779869 & 0.0122 & 0.08 \\
IA-1 & 80.132064 & -60.666987 & 0.0359 &  &  & 0.0943 & \\
IA-2 & \multicolumn{2}{c}{$^c$} & 0.1647 & \multicolumn{2}{c}{$^c$} & 0.4329 & \\
IA-3 & 80.320806 & -60.683301 & 0.0534 & \multicolumn{2}{c}{$^c$} & 0.1403 & \\
IA-3 & 80.319622 & -60.683301 & 0.0534 & \multicolumn{2}{c}{$^c$} & 0.1403 & \\
IA-4 & \multicolumn{2}{c}{$^c$} & 0.2537 & \multicolumn{2}{c}{$^c$} & 0.6665 & \\ 
\enddata
\\[-4ex] 
 \tablecomments{\\
$^a$  Average number of accidental coincidences expected in the region searched\\
$^b$  Difference between USNO and 2MASS coordinates of candidate counterparts\\
$^c$  Not searched because of non-negligible probability for accidental coincidence $N_{r99}$\\}
 \end{deluxetable}

\clearpage

\begin{deluxetable}{lcccccccccccc}
\tabletypesize{\small}
  \tablewidth{0pc}
\tablecaption{Near-infrared magnitudes and colors of candidate 2MASS counterparts. \label{t:2MASS_colors}}
  \tablehead{(1) & (2) & (3) & (4) & (5) & (6) & (7)}
  \startdata
Source & J & H & K$_S$ & J-H & H-K$_S$ & J-K$_S$ \\[0.5ex] \hline\\[-2.5ex]
S3-5 & $16.30\pm 0.12$ & $15.71\pm 0.13$ & $14.53\pm 0.09$ & $0.59\pm 0.17$ & $1.18\pm 0.16$ & $1.77\pm 0.14$ \\
S3-9 & $16.24\pm 0.14$ & $15.67\pm 0.15$ & $14.61\pm 0.11$ & $0.57\pm 0.20$ & $1.05\pm 0.19$ & $1.62\pm 0.17$ \\
S3-10 & $6.83\pm 0.02$ & $6.83\pm 0.04$ & $6.79\pm 0.02$ & $0.00\pm 0.05$ & $0.04\pm 0.04$ & $0.04\pm 0.03$ \\
 \enddata
 \end{deluxetable}

\begin{deluxetable}{lccccc}
  \tablewidth{0pc}
\tablecaption{Fits to X-ray spectra of the two brightest \chandra\ sources. \label{t:spectra}}
  \tablehead{(1) & (2) & (3) & (4) & (5) & (6)}
  \startdata
Source &  Model & \chisq & $\nu$ & $N_H/(10^{22}\, {\rm cm}^{-2})$ & $\Gamma$ or $kT/(\kev)$  \\[0.5ex]
\hline\\[-2.5ex]
S3-5  & {\tt powerlaw} & 52.9 & 59 & 0.0 (0.00--0.03)$^a$  & 2.05 (1.97--2.16)$^a$\\
S3-5  & {\tt mekal}    & 120  & 59 & $^b$ &   $^b$  \\
S3-5  & {\tt bbody}    & 232  & 59 & $^b$ &   $^b$  \\
S3-10 & {\tt powerlaw} & 10.1 & 4  & $^b$ &   $^b$  \\
S3-10 & {\tt mekal}    & 5.7  & 4  & 0.0  (0.00--0.37)$^a$ & 0.42 (0.28--0.51)$^a$ \\
S3-10 & {\tt bbody}    & 5.5  & 4  & 1.27 (1.09--1.45)$^a$ & 0.07 (0.05--0.10)$^a$ \\
 \enddata
\\[-4ex] 
\tablecomments{\\
$^a$  68\%-confidence intervals for two interesting parameters\\
$^b$  No parameter determined for statistically unacceptable fit\\
 }
 \end{deluxetable}
 
\begin{deluxetable}{lccccccc}
\tabletypesize{\small}
  \tablewidth{0pc}
\tablecaption{Emission lines in the EMMI spectrum of Leon X-1. \label{t:emmi-lines}}
  \tablehead{(1) & (2) & (3) & (4) & (5) & (6) & (7) & (8)}
  \startdata
Emission & Rest \l & Obs.\ \l &   $z$  &       FWHM$^b$       &  Flux$^c$   &  $F/F($\hb$_N)$        & $F/F($\hb$_B)$\\
line$^a$ &  (\AA)  &  (\AA)   &        &     (km s$^{-1}$)    & ($F_{15}$)  &             $^d$        &      $^d$     \\ \hline \\[-2ex]
\nev     &  3426   &  4520.6  & 0.3195 &  $719_{-182}^{+243}$ &    $1.21_{-0.45}^{+0.74}$ & $0.33_{-0.14}^{+0.26}$ & $0.12_{-0.05}^{+0.10}$ \\
\oii     &  3727   &  4924.5  & 0.3213 &   $598_{-92}^{+96}$  &    $0.48_{-0.12}^{+0.14}$ & $0.13_{-0.04}^{+0.06}$ & $0.05_{-0.02}^{+0.02}$ \\
\fevii   &  3760   &  4961.1  & 0.3194 & $1021_{-150}^{+172}$ &    $0.57_{-0.15}^{+0.18}$ & $0.15_{-0.05}^{+0.07}$ & $0.06_{-0.02}^{+0.03}$ \\
\neiii   &  3869   &  5106.4  & 0.3198 &   $712_{-89}^{+95}$  &    $0.57_{-0.12}^{+0.13}$ & $0.15_{-0.04}^{+0.06}$ & $0.06_{-0.02}^{+0.02}$ \\
\hei+\hz &  3889   &  5135.5  & 0.3205 & $1175_{-146}^{+157}$ &    $0.71_{-0.15}^{+0.17}$ & $0.19_{-0.06}^{+0.07}$ & $0.07_{-0.02}^{+0.03}$ \\
\he      &  3970   &  5241.1  & 0.3202 & $1419_{-172}^{+170}$ &    $0.96_{-0.21}^{+0.23}$ & $0.26_{-0.08}^{+0.10}$ & $0.10_{-0.03}^{+0.04}$ \\
\hd$_N$  &  4102   &  5417.4  & 0.3207 & $1222_{-238}^{+258}$ &    $0.66_{-0.24}^{+0.29}$ & $0.18_{-0.08}^{+0.11}$ & $0.07_{-0.03}^{+0.04}$ \\
\hd$_B$  &  4102   &  5417.7  & 0.3207 & $4681_{-614}^{+949}$ &    $3.24_{-0.79}^{+1.13}$ & $0.88_{-0.28}^{+0.43}$ & $0.33_{-0.11}^{+0.17}$ \\
\hg$_N$  &  4342   &  5731.4  & 0.3200 & $1278_{-172}^{+170}$ &    $1.45_{-0.34}^{+0.37}$ & $0.40_{-0.12}^{+0.15}$ & $0.15_{-0.05}^{+0.06}$ \\
\hg$_B$  &  4342   &  5732.6  & 0.3203 & $4909_{-372}^{+440}$ &    $3.98_{-0.78}^{+0.94}$ & 1$.08_{-0.29}^{+0.40}$ & $0.41_{-0.11}^{+0.15}$ \\
\oiii    &  4363   &  5757.4  & 0.3196 &  $905_{-181}^{+236}$ &    $0.56_{-0.19}^{+0.27}$ & $0.15_{-0.06}^{+0.10}$ & $0.06_{-0.02}^{+0.04}$ \\
\hb$_N$  &  4861   &  6420.4  & 0.3208 &  $1343_{-62}^{+63}$  &    $3.67_{-0.36}^{+0.37}$ & $1.00_{-0.18}^{+0.22}$ & $0.37_{-0.07}^{+0.09}$ \\
\hb$_B$  &  4861   &  6422.2  & 0.3212 & $4481_{-194}^{+212}$ &    $9.80_{-1.02}^{+1.13}$ & $2.67_{-0.50}^{+0.62}$ & $1.00_{-0.20}^{+0.24}$ \\
\oiii    &  4959   &  6547.0  & 0.3202 &   $786_{-61}^{+64}$  &    $1.10_{-0.15}^{+0.16}$ & $0.30_{-0.06}^{+0.08}$ & $0.11_{-0.03}^{+0.03}$ \\
\oiii    &  5007   &  6610.6  & 0.3203 &   $501_{-22}^{+22}$  &    $1.76_{-0.13}^{+0.14}$ & $0.48_{-0.08}^{+0.09}$ & $0.18_{-0.03}^{+0.04}$ \\
\fevii   &  5721   &  7546.6  & 0.3191 &  $744_{-234}^{+220}$ &    $0.19_{-0.09}^{+0.12}$ & $0.05_{-0.03}^{+0.04}$ & $0.02_{-0.01}^{+0.02}$ \\
\hei     &  5876   &  7761.0  & 0.3208 & $2935_{-278}^{+299}$ &    $1.20_{-0.20}^{+0.23}$ & $0.33_{-0.08}^{+0.10}$ & $0.12_{-0.03}^{+0.04}$ \\
\ha$_N$  &  6563   &  8669.4  & 0.3210 &  $1513_{-27}^{+26}$  &   $13.93_{-0.61}^{+0.60}$& $3.79_{-0.50}^{+0.58}$ & $1.42_{-0.20}^{+0.23}$ \\
\ha$_B$  &  6563   &  8670.4  & 0.3211 & $4053_{-119}^{+125}$ &   $15.29_{-1.38}^{+1.50}$& $4.16_{-0.73}^{+0.89}$ & $1.56_{-0.29}^{+0.35}$ \\
 \enddata
\\[-4ex] 
 \tablecomments{\\
$^a$  Broad component of double gaussian fit denoted by subscript $B$; narrow, by subscript $N$\\
$^b$  Rest-frame full width at half-maximum, corrected for instrumental resolution\\
$^c$  Flux in units of $10^{-15}$ erg s$^{-1}$ cm$^{-2}$\\
$^d$  Propagated statistical fitting error combined with estimated systematic uncertainty\\
}
 \end{deluxetable}
 
\begin{deluxetable}{lccccccc}
\tabletypesize{\small}
  \tablewidth{0pc}
\tablecaption{Emission lines in the EFOSC2 spectrum of Leon X-1.\label{t:efosc2-lines}}
  \tablehead{(1) & (2) & (3) & (4) & (5) & (6) & (7) & (8)}
  \startdata
Emission & Rest \l & Obs.\ \l &   $z$  &       FWHM$^b$       &  Flux$^c$   &  $F/F($\hb$_N)$        & $F/F($\hb$_B)$\\
line$^a$ &  (\AA)  &  (\AA)   &        &     (km s$^{-1}$)    & ($F_{15}$)  &             $^d$        &      $^d$     \\ \hline \\[-2ex]
\hb$_N$  &  4861   &  6418.1  & 0.3203 &   $1236_{-40}^{+44}$ &    $3.65_{-0.27}^{+0.31}$ & $1.00_{-0.15}^{+0.17}$ & $0.30_{-0.04}^{+0.05}$ \\
\hb$_B$  &  4861   &  6423.3  & 0.3214 &  $3757_{-90}^{+125}$ &   $12.22_{-0.75}^{+0.96}$& $3.35_{-0.46}^{+0.55}$ & $1.00_{-0.13}^{+0.15}$ \\
\oiii    &  4959   &  6548.9  & 0.3206 &    $621_{-22}^{+23}$ &    $1.25_{-0.08}^{+0.09}$ & $0.34_{-0.05}^{+0.05}$ & $0.10_{-0.01}^{+0.01}$ \\
\oiii    &  5007   &  6612.3  & 0.3206 &    $528_{-7}^{+7}$   &    $2.59_{-0.07}^{+0.07}$ & $0.71_{-0.07}^{+0.08}$ & $0.21_{-0.02}^{+0.02}$ \\
\fevii   &  5721   &  7550.0  & 0.3197 &    $477_{-59}^{+66}$ &    $0.16_{-0.04}^{+0.04}$ & $0.05_{-0.01}^{+0.02}$ & $0.01_{-0.00}^{+0.00}$ \\
\hei$_N$ &  5876   &  7757.1  & 0.3201 &   $537_{-98}^{+111}$ &   $0.15_{-0.05}^{+0.06}$ & $0.04_{-0.02}^{+0.02}$ & $0.01_{-0.00}^{+0.01}$ \\
\hei$_B$ &  5876   &  7760.8  & 0.3208 & $2587_{-118}^{+134}$ &    $2.10_{-0.19}^{+0.21}$ & $0.58_{-0.09}^{+0.11}$ & $0.17_{-0.03}^{+0.03}$ \\
\fevii   &  6087   &  8031.7  & 0.3195 &    $482_{-32}^{+34}$ &    $0.32_{-0.04}^{+0.04}$ & $0.09_{-0.02}^{+0.02}$ & $0.03_{-0.00}^{+0.01}$ \\
 \enddata
\\[-4ex] 
 \tablecomments{\\
$^a$  Broad component of double gaussian fit denoted by subscript $B$; narrow, by subscript $N$\\
$^b$  Rest-frame full width at half-maximum, corrected for instrumental resolution\\
$^c$  Flux in units of $10^{-15}$ erg s$^{-1}$ cm$^{-2}$\\
$^d$  Propagated statistical fitting error combined with estimated systematic uncertainty\\
}
 \end{deluxetable}

\end{document}